%% file: main.tex
\documentclass[sigconf]{acmart}
\usepackage{mathrsfs,mathtools}
\usepackage{algorithm}
\usepackage{epsfig}
\usepackage{xcolor,colortbl,flushend}
\usepackage{subfigure}
\usepackage{verbatim}
\usepackage{enumerate}
\usepackage{url}
\usepackage{bm,bbm}
\usepackage[font=small,skip=0pt]{caption}
\usepackage{color}
\usepackage{soul}
\usepackage[noend]{algpseudocode}
\usepackage{graphicx}
\usepackage[figuresright]{rotating}
\usepackage{geometry}
\usepackage{multirow}
\usepackage{booktabs}
\usepackage{appendix}
\usepackage{tcolorbox}

\copyrightyear{2024}
\acmYear{2024}
\setcopyright{rightsretained}
\acmConference[WWW '24 Companion]{Companion Proceedings of the ACM Web Conference 2024}{May 13--17, 2024}{Singapore, Singapore}
\acmBooktitle{Companion Proceedings of the ACM Web Conference 2024 (WWW '24 Companion), May 13--17, 2024, Singapore, Singapore}\acmDOI{10.1145/3589335.3651257}
\acmISBN{979-8-4007-0172-6/24/05}
\begin{document}

\pagestyle{empty}

\title{When Crypto Economics Meet Graph Analytics and Learning}

\author{Bingqiao Luo}
\affiliation{%
  \institution{Supervised by Professor Bingsheng He\\
  National University of Singapore} 
  \country{Singapore}
}
\email{luo.bingqiao@u.nus.edu}

\begin{abstract}
Utilizing graph analytics and learning has proven to be an effective method for exploring aspects of crypto economics such as network effects, decentralization, tokenomics, and fraud detection. However, the majority of existing research predominantly focuses on leading cryptocurrencies, namely Bitcoin (BTC) and Ethereum (ETH), overlooking the vast diversity among the more than 10,000 cryptocurrency projects. This oversight may result in skewed insights. In our paper, we aim to broaden the scope of investigation to encompass the entire spectrum of cryptocurrencies, examining various coins across their entire life cycles. Furthermore, we intend to pioneer advanced methodologies, including graph transfer learning and the innovative concept of ``graph of graphs''. By extending our research beyond the confines of BTC and ETH, our goal is to enhance the depth of our understanding of crypto economics and to advance the development of more intricate graph-based techniques.

\end{abstract}



\maketitle

\input{sections/intro.tex}

\input{sections/problem.tex}

\input{sections/state-of-the-art.tex}

\input{sections/approach.tex}

\input{sections/methodology.tex}

\input{sections/results.tex}

\input{sections/conclusions-and-future-work.tex}

\begin{acks}
This research is supported by the National Research Foundation, Singapore under its Industry Alignment Fund – Pre-positioning (IAF-PP) Funding Initiative. Any opinions, findings and conclusions or recommendations expressed in this material are those of the author(s) and do not reflect the views of National Research Foundation, Singapore.
\end{acks}

\bibliographystyle{ACM-Reference-Format}
\bibliography{reference}

\end{document}

%% file: sections/intro.tex
\section{Introduction}

The rapid growth of Web3 and crypto markets, supported by blockchain technology, marks a significant shift in the digital economy towards enhanced transparency and accessibility. Cryptocurrencies like Bitcoin (BTC) and Ethereum (ETH) have led to the development of innovative financial services and applications, including decentralized finance (DeFi) and non-fungible tokens (NFTs). Furthermore, the rise of Web3, as the next evolution of the internet, is attracting considerable interest and investment. Crypto economics plays a crucial role in shaping the Web3 landscape and the wider cryptocurrency markets by merging blockchain technology, network theory, and economic incentives to create decentralized systems. These systems allow for a range of applications, each governed by its own economic rules and incentives, which ensure network integrity, incentivize participation, and facilitate transparent transactions.

The landscape of graph analytics and learning in crypto markets encompasses a wide range of research areas, such as transaction network analysis, fraud detection, market behavior, decentralization, and tokenomics. Studies examine the inter-correlation and influence of major cryptocurrencies \cite{ren2020tail}. Research also delves into decentralization, evaluating blockchain network topologies for robustness and efficiency \cite{cheng2021decentralization}, and investigates tokenomics by examining token distribution and incentives \cite{chen2020traveling}. 

These studies enhance the understanding of crypto markets but most existing studies only focus on one or two major cryptocurrencies like Bitcoin (BTC) or Ethereum (ETH) \cite{ferretti2020ethereum, li2022ttagn}. This restricted scope overlooks the vast and diverse landscape of over 10,000 crypto assets and introduces several limitations to our understanding of the cryptocurrency market:

Firstly, while BTC and ETH are undoubtedly market leaders, they are also among the older entities in this fast-evolving domain. Consequently, they may not fully encapsulate the nuances of more recent developments, such as NFTs and other emerging trends.

Secondly, by concentrating only on BTC and ETH, these studies fail to represent the broader spectrum of the cryptocurrency ecosystem. For instance, using BTC and ETH as proxies for understanding the project lifecycle of various cryptocurrencies can lead to skewed or incomplete insights, as they might not mirror the dynamics of newer or less dominant projects.

Lastly, since BTC and ETH may not fully represent the entire cryptocurrency market, conclusions drawn from studies focusing exclusively on them could lead to a misrepresentation of the overall crypto economics.

Therefore, it is crucial to adopt a more inclusive research approach that encompasses the entire market. This means studying a variety of coins and projects, each with their unique life cycles and characteristics. In our thesis, we aim to explore the profound impact of graph processing on various economic concepts within the crypto world. Unlike existing research that often limits its focus to major cryptocurrencies like BTC or ETH, our approach is more comprehensive. The interplay between crypto economics and the idea of ``graph of graphs'' have opened many challenges and opportunities for future research. First, we examine the \textbf{network effects} in the life cycles of crypto assets by analyzing thousands of projects from their inception to decline, identifying key factors for success. Second, we delve into \textbf{decentralization}, studying structures of various transaction graphs within and across multiple blockchains to understand transaction causes and implications. Third, we explore \textbf{tokenomics}, categorizing different models to shed light on the economic mechanisms that drive digital assets and influence project outcomes. Lastly, we investigate \textbf{risk and fraud detection} across blockchain environments, employing methods like transfer learning to leverage the varying data availability, such as Ethereum's comprehensive dataset.

In summary, our thesis seeks to utilize graph processing not just at a surface level, but in a multi-dimensional manner that allows us to understand the complexities of crypto economics more thoroughly. This comprehensive approach goes beyond existing work, offering new insights into the dynamic crypto world.

%% file: sections/problem.tex
\section{Problem} 
The primary problem addressed in this paper is the gap in understanding and effectively leveraging the interplay between crypto economics and graph learning and analytics across the entire crypto world. There is currently a lack of comprehensive understanding of crypto economics, coupled with limited development in graph analytics and learning in the crypto market, which primarily focuses on major cryptocurrencies \cite{ren2020tail, nguyen2023volatility}. We detail the challenge of uncovering this interplay in two main aspects:

\textbf{From Economics to Graph Learning \& Analytics.} The potential of applying crypto economic principles to enhance Graph Learning \& Analytics methodologies remains underexplored. For example, by incorporating life cycle theory into graph models to analyze crypto assets from inception, growth, maturity, to decline, we anticipate a deeper insight into graph dynamics and implications. This strategy may highlight the need for tailored graph learning approaches across different stages of crypto assets. Furthermore, it could foster the creation of novel graph-related metrics to understand market dynamics and project evolution, facilitating the comparative analysis and categorization of various crypto assets.

\textbf{From Graph Learning \& Analytics to Economics.} The application of Graph Learning \& Analytics in studying the economic interactions within the crypto world remains narrowly focused, with only a few studies targeting major cryptocurrencies. This oversight leaves vast potential unexplored. For example, by analyzing the graph properties of diverse crypto assets, we aim to extract valuable insights from their successes and failures, categorizing them to inform the future development of tokenomics. Additionally, studying the overall graph structure and the correlations between nodes will assist in implementing appropriate regulatory methods, contributing to a healthier and sustainable crypto market.



%% file: sections/state-of-the-art.tex
\section{State of the art}
The current work are mostly focusing on one direction of the interplay: "From Economics to Graph Learning \& Analytics". In the following, we review the related work about graph learning and analytics in four aspects in crypto economics: network effects, decentralization, tokenomics, and fraud detection.

\textbf{Network effects:} The network effect, illustrating how a product's value grows with its user base, has been analyzed in crypto economics using graph analytics, focusing mainly on major cryptocurrencies. Ren et al. \cite{ren2020tail} investigated the 30 largest cryptocurrencies from 2019 to 2020 by creating a network of cryptocurrencies linked by regression coefficients of returns, finding significant inter-cryptocurrency influence, particularly during the COVID-19 crisis. Nguyen et al. \cite{nguyen2023volatility} expanded this analysis to include 34 cryptocurrencies from 2019 to 2021. 
Their findings reveal that the network structures reflect investor decisions. Broadening the scope, Vasan et al. \cite{vasan2022quantifying} studied 50,723 NFTs minted on Foundation before June 2021 and underscored the network effect's significant role in determining NFT values among artists and collectors. 

\textbf{Decentralization:} The comparison between decentralized and centralized markets, an ongoing subject of research, has gained attention with the rise of decentralized finance. With the development of decentralized finance, growing studies focused on the evaluation of decentralization on crypto markets using graph analysis. In a notable study, Cheng et al. \cite{cheng2021decentralization} analyzed Bitcoin's network from its beginning to May 2020, identifying three phases with distinct decentralization patterns and noting higher asset centralization in the top 100 addresses. Zhang et al. \cite{zhang2023blockchain} investigated decentralization in four decentralized banks, uncovering their key external transaction core addresses. Ao et al. \cite{ao2023decentralized} investigated the AAVE token network, revealing a core-periphery structure. However, comprehensive research across all crypto assets is still lacking.

\textbf{Tokenomics:} Many studies have built network features to study the crypto economics. They mainly focused on Bitcoin and Ethereum, revealing evolving network patterns and structures, such as graph size and average degree, that correlate with economic indicators like price and volatility, suggesting their potential predictive value for cryptocurrency markets \cite{kondor2014rich, motamed2019quantitative}. 
However, research extending to the broader market is scarce. Chen et al. \cite{chen2020traveling} explored the ERC20 token ecosystem, finding many inactive tokens and signs of a nascent token economy, but focused only on its initial stages. 

\textbf{Fraud detection:} Graph-based techniques are increasingly employed in crypto economics for risk and fraud detection. Utilizing advanced graph methodologies like node and graph embedding, and Graph Neural Network (GNN) models \cite{zhou2020graph}, these techniques have significantly improved fraud detection in crypto markets \cite{li2022ttagn, weber2019anti}. Major frauds identified include Ponzi schemes in Ethereum contracts \cite{bartoletti2020dissecting}, Ethereum phishing address identification \cite{li2022ttagn}, and Bitcoin money laundering detection \cite{weber2019anti}. The use of heterogeneous graphs is notable in handling diverse node types, such as wallets and contracts \cite{jin2022heterogeneous}. However, progress in this field hinges on the availability of accurate labels, with current emphasis largely Bitcoin and Ethereum. This suggests potential for broader research.

%% file: sections/approach.tex
\section{Proposed approach}
While previous studies \cite{cheng2021decentralization, li2022ttagn} have shown the effectiveness of graph analytics and learning in the context of cryptocurrencies, their focus has predominantly been limited to BTC and ETH. This restricted scope introduces several limitations to our understanding of the cryptocurrency market, as outlined in Introduction. We propose to  develop a comprehensive understanding of the interplay between crypto economics and graph learning in the entire crypto universe.

\textbf{Technical Challenges.} Our research focuses on innovating graph analytics and learning in crypto economics, targeting three primary challenges. First, the notable lack of labels for new crypto projects and blockchains prompts us to utilize graph transfer learning \cite{lee2017transfer} to overcome data scarcity, leveraging insights from established cryptocurrencies like Ethereum. Second, the rapid dynamics and high transaction volumes of transaction graphs require a departure from traditional analysis methods aimed at simpler graphs, necessitating novel analysis and classification approaches for large-scale transaction graphs. Third, the diversity of crypto addresses, including various kinds of wallets and contracts, introduces node heterogeneity, for which we aim to develop advanced heterogeneous graph learning techniques to effectively manage this complexity.

\textbf{Proposed Research:} The following technologies enable this line of research:

First, \textbf{spatial and temporal analyses} are conducted on graphs of multiple crypto assets within and across different blockchains, such as Ethereum and Solana, to examine their structural characteristics and temporal behaviors. This investigation, which will encompass over 10,000 crypto projects, aims to compare and understand their life cycles and tokenomics in depth.

Second, \textbf{graph transfer learning}, an advanced technique in graph learning, is explored with a particular focus. This technique tackles the challenge of transferring knowledge and patterns gleaned from data-rich graphs, such as Bitcoin or Ethereum, to those cryptocurrency graphs that are data-scarce \cite{Zhang2024CollaborateTA}. It requires adjusting data formats and accounting for varied asset adoption traits for effective knowledge transfer.

Third, \textbf{``graph of graphs''}, as an innovative method, are introduced. This concept envisages each individual transaction graph as a node within a larger, interconnected network that collectively represents the entire crypto universe. This approach is aimed at deepening our understanding of the interrelationships among various cryptocurrency projects and the overarching market dynamics. This method has broad applications, such as identifying trends via transaction analysis across different projects.

%% file: sections/methodology.tex
\section{Methodology}

\begin{figure}
\centering 
\includegraphics[width=1\linewidth]{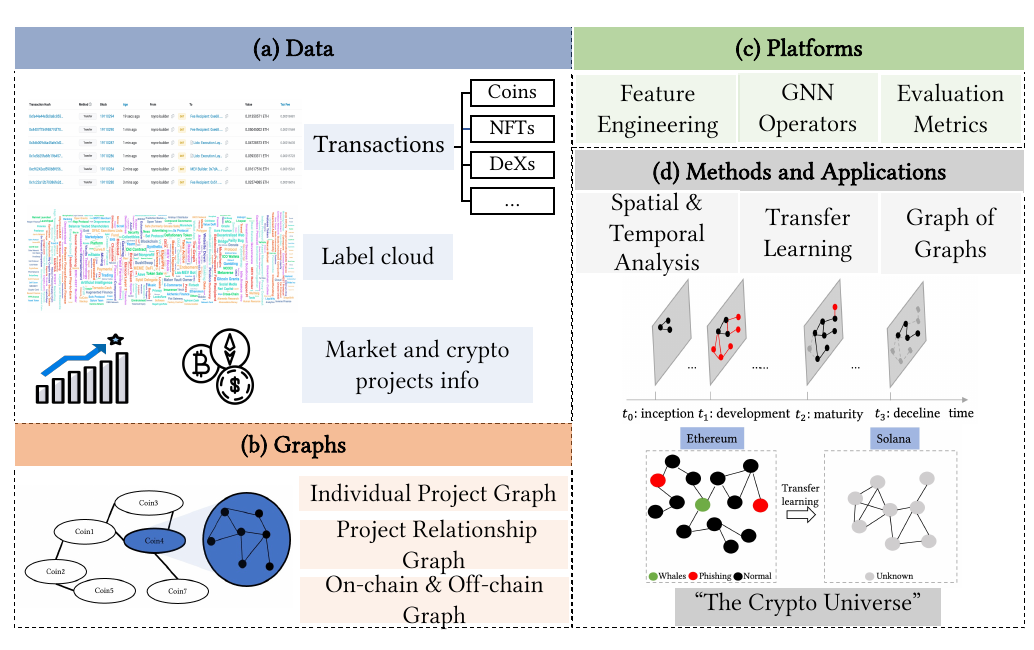} 
\caption{Overview of Methodology} 
\label{fig:methodology} 
\vspace{-2em}
\end{figure}

\autoref{fig:methodology} shows an overview of our methodology, which involves data collection and filtering, graph construction, platform development, and models and applications deployment.

\textbf{(a) Data} We collect transaction data by synchronizing a node of blockchain or through online data, such as Google Cloud BigQuery\footnote{\url{https://cloud.google.com/bigquery/public-data}. Accessed on 30 January, 2024.}. This dataset can be filtered to identify specific transactions related to various crypto projects, including particular coins and NFTs. We also collect available labels for addresses and transactions, utilizing resources like the Etherscan label cloud\footnote{\url{https://etherscan.io/labelcloud}. Accessed on 31 January, 2024.}. Moreover, our data collection extends to market-related information, encompassing aspects like price fluctuations and intrinsic details about crypto assets, including insights from project whitepapers and social media.

\textbf{(b) Graphs} We create graphs that represent transactions and broader market dynamics. For each cryptocurrency or NFT collection, we construct transaction graphs using addresses as nodes and transactions as edges, charting the market activities of these assets. Additionally, we develop project relationship graphs for a comprehensive view of the market. In these graphs, cryptocurrencies or projects act as nodes, with their correlations serving as edges. Each node represents the transaction activities of an asset, which can also be represented by a graph. Furthermore, we develop graph-to-graph matching to connect on-chain and off-chain activities, linking crypto and social media activities to enhance our understanding of the crypto ecosystem.

\textbf{(c) Platforms} we develop a specialized platform to streamline data collection and graph construction processes. This platform will not only facilitate efficient feature engineering for subsequent applications but also support diverse GNN operators. We aim to establish a standardized framework for multiple evaluation metrics, ensuring a consistent and fair assessment of our models.

\textbf{(d) Methods and Applications} we incorporate diverse techniques, including cross-chain spatial and temporal analysis, transfer learning, and ``graph of graphs'' approach. For transfer learning, the evaluation will focus on the model's efficacy in areas like fraud detection and user profiling, assessing how well insights from one domain can be applied to another. For spatial and temporal analysis, our evaluation strategy will center on the model's ability to accurately interpret and predict market trends over time and across different market segments. As for the ``graph of graphs'' approach, we aim to evaluate its capacity in synthesizing and interpreting complex, multi-layered data structures, offering a more nuanced and interconnected view of the blockchain ecosystem. This evaluation strategy aims to evaluate the adaptability, precision, and overall efficacy of our proposed approaches in dealing with the complex and dynamic nature of crypto markets.

%% file: sections/results.tex
\section{Preliminary Results} 
In our paper \cite{zhang2023live}, we gathered all ERC721 NFT transaction data on the Ethereum blockchain, created a live graph lab of the NFT transaction network, and analyzed its dynamics. We are continually updating our dataset with the most recent transactions. Additionally, in our paper \cite{wang2023etgraph}, we linked Ethereum with X (formerly Twitter) to enhance the integration of social network data into blockchain analysis. With these complete and continuously evolving datasets, our subsequent study delved into the life cycles of crypto assets, starting with NFT collections. In the following, we present the preliminary results. 

\textbf{Experimental Setup.}  Given that many NFT collections have either a very short lifespan or minimal transactions, we chose 510 NFT collections with at least 10,000 sales as representative samples for our analysis. We constructed daily \emph{MultiDiGraphs} for the transactions of each NFT collection, using addresses as nodes and transactions as edges. Note, a MultiDiGraph is a directed graph that can include multiple edges between two nodes. We then explored various potential graph metrics to understand the life cycles of NFTs, specifically looking at the number of edges, the weighted edge (by transaction volume), and the number of nodes.

\noindent
\begin{tcolorbox}[width=\linewidth,colback=white,boxrule=1pt,arc=0pt,outer arc=0pt,left=0pt,right=0pt,top=0pt,bottom=0pt,boxsep=1pt,halign=left]
{\textbf{Observation (1):}  Among various metrics, the number of edges offers a more nuanced understanding of NFT project life cycles, better reflecting the actual transactional activity. }
\end{tcolorbox}
In comparison with the number of edges, although transaction volume (weighted edges) is a common metric in financial analyses, it may not accurately reflect the life cycles of NFT collections. For instance, in collections dominated by low-price items, a single sale of a high-price item can largely influence the overall volume, obscuring the collection's typical transaction scale. Another metric we studied is the number of nodes. This metric measures the number of market participants but overlooks the depth of their engagement. Significantly, accounts that engage in multiple transactions in a single day can skew market activity perceptions due to their pronounced involvement.

\begin{figure}
\centering 
\includegraphics[width=0.9\linewidth]{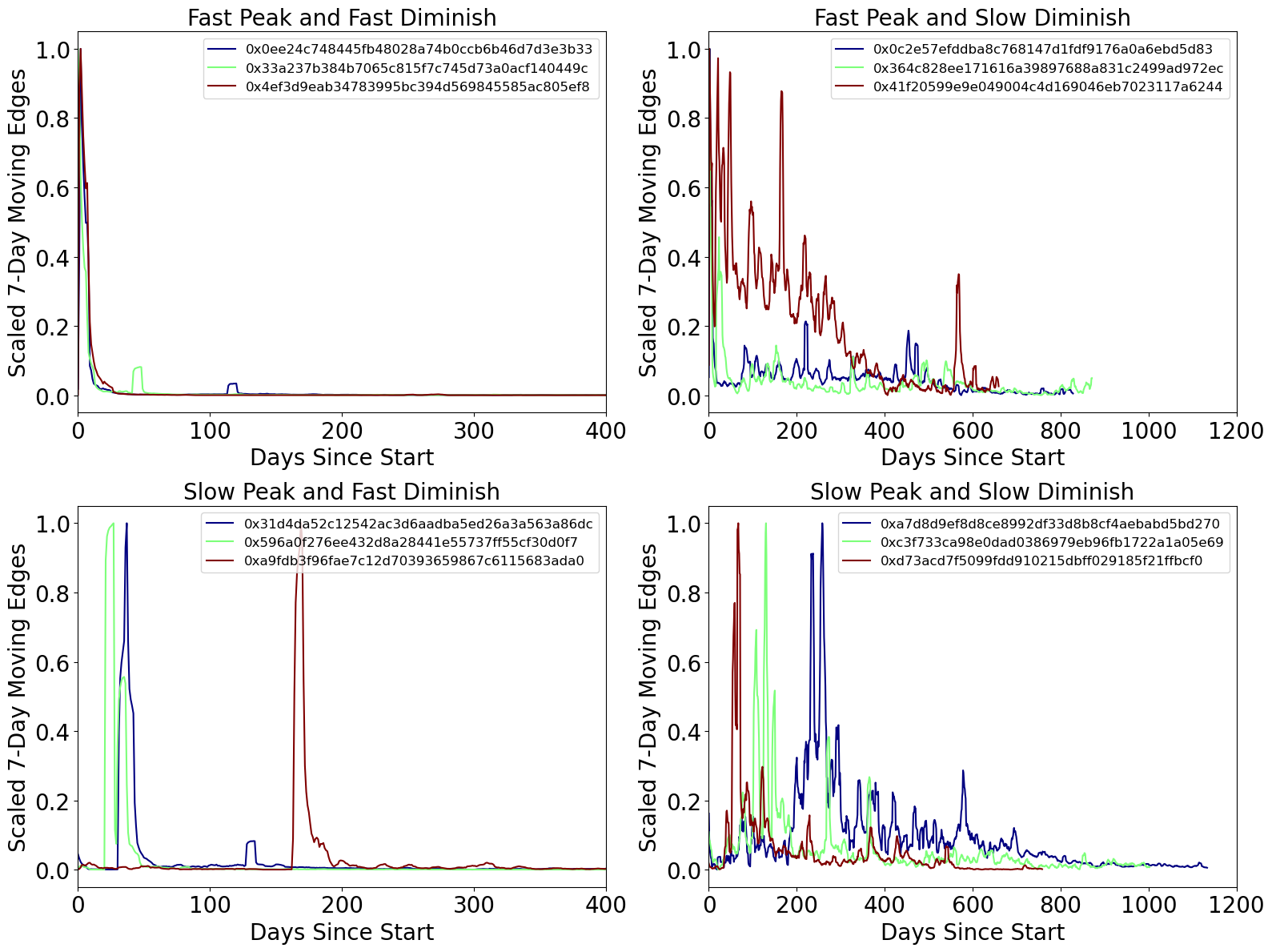}
\caption{Life Cycles of NFT Projects} 
\label{fig:lifespan} 
\vspace{-1.7em}
\end{figure}

\noindent
\begin{tcolorbox}[width=\linewidth,colback=white,boxrule=1pt,arc=0pt,outer arc=0pt,left=0pt,right=0pt,top=0pt,bottom=0pt,boxsep=1pt,halign=left]
{\textbf{Observation (2):}  Life cycles of NFT collections follow very sharp and burst patterns. }
\end{tcolorbox}
Our analysis focuses on delineating the life cycle of NFT collections through the examination of two pivotal time points: 1) \(\mathbf{t_{peak}}\): The interval from a project's launch to the peak of daily transactional activity (\(n_{peak}\)); 2) \(\mathbf{t_{diminishing}}\): The first day after \(\mathbf{t_{peak}}\) when transactional activity declines to or below 1\% of \(n_{peak}\).

\autoref{fig:lifespan} illustrates the lifespan of various NFT projects, classified into four categories based on their time from start to peak (\(t_{\mathit{start\_to\_peak}}\)) and time from peak to diminishing (\(t_{\mathit{peak\_to\_diminishing}}\)). Through this approach, we discovered that the average period from the project's start to its peak is approximately 17 days, and the average duration from peak to diminishing is about 103 days. However, a closer examination of individual collections reveals that roughly 63.3\% of them reach their peak within just two days, indicating a highly skewed distribution for \(t_{\mathit{start\_to\_peak}}\).



%% file: sections/conclusions-and-future-work.tex
\section{Conclusions and future work}
This paper presents a novel approach to examining the interplay between crypto economics, graph analytics and learning. By leveraging advanced graph techniques, such as graph transfer learning and the concept of graph of graphs, our goal is to explore and comprehend the economic mechanisms governing the entire crypto universe. Our on-going and future work is to address the technical challenges and realize the enabling technologies in the thesis.